\documentclass[rnote]{aa}  

\usepackage{graphicx}
\usepackage{txfonts}
\begin{document}

   \title{Spot modelling of periodic weak-line T~Tauri stars observed by CoRoT in NGC~2264 }
              \titlerunning{Spot models of WTTS in NGC~2264}
  \authorrunning{A. F. Lanza et al.}


   \author{A.~F.~Lanza\inst{1}
          \and E.~Flaccomio\inst{2}
          \and   S.~Messina\inst{1}
          \and G.~Micela\inst{2}   
          \and I.~Pagano\inst{1}
          \and G.~Leto\inst{1}
          }

   \institute{INAF-Osservatorio Astrofisico di Catania, Via S.~Sofia, 78 - 95123 Catania, Italy\\
         \email{nuccio.lanza@oact.inaf.it, sergio.messina@oact.inaf.it, isabella.pagano@oact.inaf.it, giuseppe.leto@oact.inaf.it}
         \and
             INAF-Osservatorio Astronomico di Palermo "G.~S.~Vaiana", Piazza del Parlamento, 1 -- 90134 Palermo, Italy\\
             \email{ettoref@astropa.inaf.it, giusi@astropa.inaf.it}
             }

   \date{Received ... ; accepted ...}

 
  \abstract
   {The space telescope CoRoT has provided light curves of T~Tauri stars belonging to the star-forming region of NGC~2264 with unprecedented continuity and precision in the framework of a coordinated multi-wavelength observational project.}
   {We perform  spot modelling  of the optical light curves of five weak-line T~Tauri stars whose variability is dominated by starspots.  }
   {We apply {{ two-spot and evolving single spot models }} in the framework of a Bayesian Monte Carlo Markov Chain approach to derive the a posteriori distribution of the starspot parameters and the inclination of the star rotation axis. We focus on the rotation periods of the spots that can provide evidence for differential rotation in those stars.}
   {We find meaningful results in the case of three stars with an inclination higher than $50^{\circ}$ and a slow variation of the light curve amplitude. The relative difference of the spot rotation periods ranges from $0.02$ to $0.05$ that is  $3-10$ times larger than the amplitude of the differential rotation found in similar stars with Doppler imaging techniques.}
   {We conclude that the intrinsic starspot evolution, although very slow, has a significant impact on the determination of the differential rotation by means of our spot modelling approach. We estimate typical timescales for the evolution of the starspot pattern between $\sim 20$ and $\sim 50$ rotation periods in our stars. }


   \maketitle
%

\section{Introduction}
Weak-line T~Tauri stars (hereafter WTTs), so called because of their lack of spectroscopic accrection signatures, are pre-main-sequence stars that generally display quasi-stable sinusoidal light curves attributed to cool magnetic starspots on their surface \citep[e.g.,][]{Grankinetal08}. The space telescope CoRoT \citep[Convection, Rotation, and Transits;][]{Auvergneetal09} observed a sample of such stars in the star-forming open cluster NGC~2264 in 2008 and 2011. Thanks to its relative proximity ($\sim 760$~pc), well-defined stellar population, and low foreground extinction, NGC~2264 is a primary target for star formation and pre-main-sequence star studies \citep{Dahm08}.  

CoRoT 2011 observing run was part of a campaign coordinated with several other telescopes operating in different spectral domains, including Spitzer in the infrared and Chandra in the X-rays, and  optical spectroscopy with Flames and UVES at the VLT. It was named as the Coordinated Synoptic Investigation (CSI) of NGC~2264 and its main purpose was to investigate the variability of young  stars with an unprecedented time continuity and spectral range coverage \citep[see][for details]{Codyetal14,Staufferetal14,Venutietal14,Venutietal15,McGinnisetal15}. { In the case of WTTs, the simultaneous observation in several spectral domains allows us to investigate the correlations between photospheric spots and the higher atmospheric levels in active regions both during quiescent phases and flares. This provides clues on the heating mechanisms at work in the outer atmospheres of very active, young stars and on the energy release  during flares.}

In the present work, we model the CoRoT optical light curves of five WTTs by means of a simple spot modelling technique to extract information on  spot rotation periods and their evolutionary timescales. This is particularly interesting for understanding rotation and magnetic activity in pre-main-sequence stars of solar mass. With an estimated age of its stars ranging in the $1-5$~Myr interval, NGC~2264 is an ideal laboratory to address this field of study. { The light curves of several WTTs were obtained by CoRoT with high precision  along $\sim 40$~days of uninterrupted observations. }Therefore, our exploratory investigation benefits of data of unprecedented quality that allow us to fully exploit the potentialities of spot modelling and compare our differential rotation measurements with those obtained by means of Doppler Imaging tomography on similar stars also to derive information on spot evolution in WTTs. 

\section{Observations}
CoRoT has a 27-cm aperture telescope and has been observing NGC~2264 continuously from 1$^{\rm st}$ December 2011 to 3$^{\rm rd}$ January 2012.  Targets have been observed with the CCD dedicated to  searching planetary transits. CoRoT performs aperture photometry with a pre-selected mask for each target and provides also the light curves in three non-standardized  passbands for targets brighter than $R \sim 14$ by means of a prism that disperses the star's spectrum across the mask whose pixels are read in three groups along the wavelength range according to the fraction of collected photoelectrons \citep{Auvergneetal09}. In our investigation, we always considered the total flux integrated over the whole mask of each star, i.e., by summing up the three passband fluxes in the case of  targets with $R \la 14$. These white-light curves cover the wavelength range $300-1100$~nm. They have a higher signal-to-noise ratio and are more stable than the light curves in the individual passbands.  The exposure time was  usually of 512~s, except for a subset of selected targets that had 32~s. 

\begin{figure}
   \centerline{
   \includegraphics[width=9cm,angle=0]{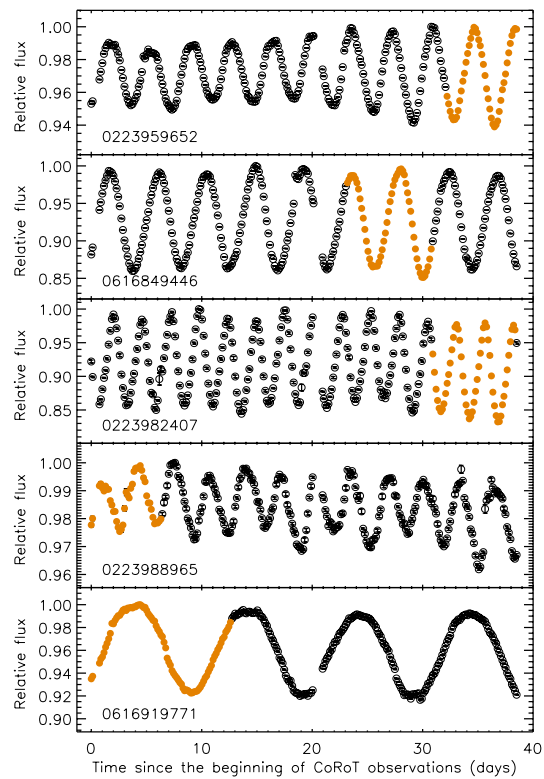}} 
   \caption{Light curves of our five target stars displaying the flux normalized to the respective maximum value vs. the time as labelled. Errorbars are within the size of the symbols and indicate the standard error of each point that is an average of individual CoRoT measurements along two consecutive satellite orbits (open black circles). Filled orange circles indicate the segment of the light curve fitted to find the spot rotation periods (see Sect.~\ref{model}). }
              \label{Fig1}
    \end{figure}

We visually inspected the light curves of several WTTs stars to find a subsample with the most stable modulation along the $\sim 40$~days of CoRoT observations and the lowest fluctuations,  indicative of low accretion and quasi-stable spots in those targets. These selection criteria make us confident to have stars whose variability is dominated by the rotational modulation produced by cool photospheric starspots. { Our sample consists of five stars listed in Table~\ref{table1} where we report, from left to right, the CoRoT ID number of the star, its Mon ID number according to \citet{Codyetal14} to allow identification  in the papers of the CSI~NGC~2264 series, the spectral type, the reference for the spectral type, the $R$ magnitude, its reference, the $I_{\rm c}$ magnitude, its reference, the equivalent width (EW) of the H$\alpha$ line, its reference, the exposure time $\tau_{\rm e}$ of CoRoT photometry, and the relative mean standard error per data point  $\epsilon$ after averaging the flux along two consecutive CoRoT orbits (see below). Most of our targets display an H$\alpha$ in absorption, i.e., a positive EW except during flares, in agreement with their WTT classification. The light curves of our stars are plotted in Fig.~\ref{Fig1}. }

The low Earth orbit of CoRoT introduces perturbations associated with the crossings of the South Atlantic Anomaly of the Earth magnetic field and with the ingress in and egress from the Earth shadow that affect the detector temperature. Most of such effects have been corrected or flagged by the reduction pipeline. In our investigation, we use  the so-called N2-level light curves that are freely available from the CoRoT Archive\footnote{doc-corot.ias.u-psud.fr} together with a detailed documentation. We discarded all the flagged flux measurements that amount to $\approx 7-15$ percent of the points along one satellite orbit and averaged the flux along two consecutive  orbits of 6184~s each to eliminate possible residual variations produced by  changes in the telescope environment. Since we are interested in starspot rotation and evolution, whose timescales are of the order of several days, our final cadence of one point per $\sim 12\, 368$~s does not introduce any limitation in our investigation and grants us an almost perfect continuity of our time series. 
\begin{table*}
\caption{Properties of WTTs observed by CoRoT and modelled in this investigation.}
\begin{tabular}{cccccccccccc}
\hline
 & & & & & &  & & \\
CoRoT ID & Mon-ID  & Sp. type & Sp.  Ref.$^{(a)}$ & $R$ & $R$ Ref$^{(a)}$ & $I_{\rm c}$ & $I_{\rm c}$ Ref.$^{(a)}$ & EW H$\alpha$ & H$\alpha$ Ref.$^{(a)}$ & $\tau_{\rm e}$ & $\epsilon$ \\ 
 &  & & & (mag) & & (mag) & & (\AA) & & (s) &  \\
 & & & & & & & & & \\
 \hline
 & & & & & & & & & \\
 223959652 &  1256  & G0  &  1   &   12.85   & 4   &  12.47  &  4  &   -3.32  &   6 & 32 & $1.432 \times 10^{-4}$ \\
 616849446 & 565 & K4.5  & 2  &  13.94   &    5    & 13.19    &   5    &  1.60  &  2  & 32 & $ 2.756 \times 10^{-4}$ \\
 223982407 & 33  & K5  &  2 &    14.47  &  4  &   13.76 &   4 &     1.40 &   2 & 512 & $1.212 \times 10^{-3}$ \\
 223988965 & 695 & K6  & 1 &     14.11 &   4  &   13.31 &    3  &    1.30 &   6 & 512 & $4.683 \times 10^{-4}$ \\
 616919771 & 784 & K5  &  1 &     14.06 &   4   &  13.33 &   4  &    1.83 &   6 &  512 & $3.488 \times 10^{-4}$ \\
 & & & & & & & & & \\
 \hline
\end{tabular}
$^{(a)}$ References: 1: \citet{Makidonetal04}; 2: \citet{DahmSimon05}; 3: \citet{Sungetal08}, Table 3 (photometry obtained at the Canada-France-Hawaii Telescope);
 4: \citet{Sungetal08}, Table 8 (SSO WFI photometry, after adding 0.067 to the I magnitude so to homogenize it with Table 3); 5: \citet{Lammetal04}; and 6: \citet{Rebulletal02}. 
\label{table1}
\end{table*}

\section{Model}
\label{model}
A requisite for measuring stellar rotation by using starspots is the stability of spots themselves, i.e, their evolution timescale must be longer than the rotation period. If we want to measure a differential rotation of amplitude $\Delta \Omega$, where $\Omega$ is the rotation angular velocity, the starspot evolution timescale should not be shorter than $1/(\Delta \Omega)$. A simple measurement of the evolution timescale of the whole spot pattern can be obtained by computing the autocorrelation of the photometric time series as discussed by, e.g., \citet{Lanzaetal14}. For stable spots, the autocorrelation function displays a periodic pattern with maxima and minima separated by a time lag $\tau$ equal to the mean rotation period $P_{\rm rot}$, while, in the case of starspots with a decaying area, the amplitude of the autocorrelation modulation decreases steadily and the rate of decrease can be used to estimate the evolution timescale. When the  maximum of the autocorrelation at $\tau = P_{\rm rot}$ is lower than 0.5-0.6 times the maximum at $\tau=0$, the evolution of the spot pattern is generally too fast  to obtain a measure of differential rotation using spot modelling in the case of solar-like stars \citep[cf.][]{Lanzaetal14}. 

A fundamental limitation is that spot modelling is an ill-posed problem because we try to reconstruct a two-dimensional map of the stellar surface starting from a photometric time series, i.e., a one-dimensional dataset. To alleviate this problem,  a priori information can be included in the process of light curve modelling to make the solution unique and stable as, e.g.,  in the case of the maximum entropy spot modelling \citep{Lanzaetal98,Lanzaetal06,Lanzaetal07}. In the case of CoRoT light curves, several approaches have been proposed taking advantage from the high precision and almost uniform sampling of the time series \citep[e.g.][]{Lanzaetal09,Frohlichetal09,Mosseretal09,Huberetal10}. 

In the present case, we apply a Monte Carlo Markov Chain (hereafter MCMC) approach to statistically sample the a posteriori parameter space and the correlation among different parameters in a  spot model. Such a technique has been proposed by \citet{Croll06} and \citet{Frohlich07} and has been applied, among others, to Kepler time series \citep{Frohlichetal12}. In principle, it allows a complete sampling of the different possible starspot configurations that fit a given light curve. In practice, the long computation time limits its applicability to rather simple models consisting of a few discrete starspots. This is a consequence of the strong correlations among the model parameters that become more and more important as the number of model starspots is increased \citep[cf. ][and references therein]{Frohlichetal12}. Specifically, if more than $2-3$ spots are assumed, the sampling of the a posteriori parameter space becomes very slow because the Markov chains cannot move freely across the space, but are constrained to follow paths determined by local parameter correlations { \citep[see][for more details]{Lanzaetal14}}. Nevertheless, a full MCMC approach is generally feasible when we consider two discrete starspots and an adjustable mean flux level to  account for the variations in a background of small spots uniformly distributed in longitude, as it was demonstrated by \citet{Lanzaetal14}. We refer to that paper for a detailed description of our approach, while we provide here only a general introduction. 

We consider an unperturbed distribution of the stellar photospheric brightness $I$ as given by a quadratic limb-darkening law:
\begin{equation}
I(\mu)/I_{0} = a_{\rm p} + b_{\rm p} \mu + c_{\rm p} \mu^{2},
\end{equation}
where $I_{0}$ is the brightness at the centre of the disc, $a_{\rm p}$, $b_{\rm p}$ and $c_{\rm p}$  the limb-darkening coefficients, and $\mu \equiv \cos \psi $, with $\psi$ the angle between the normal to a point on the photosphere and the line of sight.  For the CoRoT white passband, we adopt limb-darkening coefficients by \citet{ClaretBloemen11} (see Sect.~\ref{parameters}). 

We fit a light curve by considering a variable inclination $i$ of the stellar spin axis to the line of sight and assuming that the light variation is produced by two spots of area $A_{k}$ located at colatitude $\theta_{k}$ and longitude $\lambda_{k}$, and rotating with rotation period $P_{k}$ ($k=1,2$), in addition to a longitude-independent flux term  $F_{0}$ that can be adjusted to take into account a uniformly distributed background of small spots.  We do not consider any spot evolution  along a given fitted interval of the light curve. The spots are assumed to be point-like, that is we do not consider partial visibility of their area when they rise or set at the stellar limb. This simple model has a total of ten free parameters per light curve and their a posteriori distributions can be sampled in a reasonable time, in spite of their correlations. Only in the case of strong correlations, we need to reduce the number of free parameters, usually by fixing the inclination $i$ and/or $F_{0}$ \citep[see][for details]{Lanzaetal14}. 

The best fit to an entire light curve of $\sim 40$ days has an unacceptably large $\chi^{2}$ because of our assumption of non-evolving spots. Therefore, we cut the entire light curve into smaller and smaller equal segments  until we find acceptable best fits. For a given segment,  we compare  the model allowing for different spot rotation periods, i.e., $P_{1} \not= P_{2}$ (DR model) with that assuming rigid rotation, i.e., $P_{1} = P_{2}$ (RR model) by means of the Bayesian Information Criterion as in \citet{Lanzaetal14}. Specifically, we define the variation of the BIC statistics as 
\begin{equation}
\Delta {\rm BIC} = 2 \left( \frac{\chi^{2}_{\rm RR}}{\chi^{2}_{\rm DR}} -1 \right) -\ln N,
\end{equation}
where $\chi^{2}_{\rm RR}$ is the chi square obtained with the RR model, $\chi^{2}_{\rm DR}$ the one obtained with the DR model, and $N$ the number of data points in the fitted segment; the $-\ln N$ term accounts for the additional free parameter in the case of the DR model with respect to the RR model. The best fits of the segments are computed by means of the IDL procedure {\tt MPFIT.PRO}\footnote{http://www.physics.wisc.edu/$\sim$craigm/idl/fitting.html}, that implements a Levenberg-Marquardt minimization of the chi square with the imposed constraint of positive starspot areas, { and  the procedure {\tt AMOEBA.PRO} that implements a Nelder-Mead minimization; we assume the minimum of the two minimizations as our best fit solution.} To allow an efficient exploration of the different  light curve segments and a comparison of their best fits, the inclination has generally been fixed at $i=60^{\circ}$\footnote{This allows the code to adjust the interval of visibility of a spot by varying its latitude, that is not possible when the inclination is $90^{\circ}$ because in that case the interval of visibility is half  a rotation, independently of the latitude. In this way, the quality of the best fits with the DR and RR models is improved with respect to the $i=90^{\circ}$ case.}.  The minima found by those optimization methods may depend on their starting points in the parameter space.  The initial spot colatitudes are initially fixed at the value of the inclination, their rotation periods at the estimate given by the autocorrelation function, their areas are constrained by the depth of the light curve minimum in the segment to be fitted, and their initial longitudes, the most sensitive parameters in our model, are chosen to reproduce the phase of the observed light minimum of the fitted segment.  Then those initial longitudes are varied in steps of $90^{\circ}$ and the minimum of the $\chi^{2}$ recomputed to find the initial conditions giving the lowest minimum for each of the two optimization methods.

The DR model is considered to be statistically better than the RR model when $\Delta {\rm BIC} \geq 2$. 
 In this way, we select  segments showing evidence of differential rotation and pick up the best one by considering the largest $\Delta {\rm BIC}$ value associated with adequate best fits, that is we discard segments where the Levenberg-Marquardt or Nelder-Mead algorithms do not properly converge.  For this segment only, we perform a full MCMC Bayesian analysis as described in \citet{Lanzaetal14}.  
 
We apply the Metropolis-Hasting algorithm to define a chain of randomly distributed points in the parameter space that sample the a posteriori distributions of the model parameters \citep{Pressetal02}. If the value of the chi square corresponding to the $k$-th point of the chain is $\chi^{2}_{k}$, we accept a new point as the $(k+1)$-th of the chain if $\chi^{2}_{k+1} < \chi^{2}_{k}$, otherwise we accept it with a probability proportional to $\exp [-(\chi^{2}_{k+1} - \chi^{2}_{k})/\chi^{2}_{\rm min}]$, where $\chi^{2}_{\rm min}$ is the minimum of the chi square along the chain. To better sample the region with the highest likelihood, we start from the minimum of the chi square and consider only points that correspond to a chi square value up to 1 percent above the minimum. If a lower minimum is found by the Metropolis-Hasting algorithm, it is adopted as a new starting point for the Monte Carlo Markov Chain. The amplitude of each random step of the Metropolis-Hasting algorithm is tuned to have an average acceptance rate between 20 and 35 percent. The sampling of the a posteriori parameter space is adequate only if the distributions of the model parameters  have converged. A necessary condition for this has been expressed by Gelman \& Rubin through their $\hat{R}$ statistics that should be smaller than $1.1-1.2$ for all the parameters  \citep[cf.][]{Croll06}. To compute $\hat{R}$, we follow the scheme specified by \citet{Lanzaetal14} by splitting our long MCMCs, that consists of several million points, into four subchains that can be regarded as statistically independent and therefore can be statistically compared with each other.  

In \citet{Lanzaetal14} we performed a detailed comparison of our approach with that proposed by \citet{Croll06} in the case of $\epsilon$~Eridani and found perfectly compatible results in spite of our assumption of point-like spots and a chi square limit at 1 percent above the minimum, while Croll considered extended polar-cap spots and a somewhat larger chi square limit of $\sim 4$ percent. 

The correlations between model parameters in MCMC chains is a critical issue that has been discussed by, e.g., \citet{Croll06}, \citet{Frohlichetal12}, and \citet{Lanzaetal14}. Methods applied to other problems do not generally perform well in the case of spot modelling \citep{Lanzaetal14}. In the present case, we  treat correlations by fixing some of the correlated parameters to their best fit values as obtained from the Metropolis-Hasting optimization according to \citet{Lanzaetal14}. Future developments may consider techniques such as Principal Component Analysis. They have been applied to light curve inversion and Doppler Imaging problems considering a continuous brightness distribution \citep[][]{Berdyugina98,SavanovStrassmeier05,SavanovStrassmeier08}. However, their implementation is not straightforward when the dependence of the model on some of its parameters is non-linear as in the case of few-spot models or the noise level of the data is not precisely known as in the present case because our models cannot fit light variations on  timescales shorter than $0.2-0.3$ of the rotation period owing to their limited number of degrees of freedom (see Sect.~\ref{results}). 

The adoption  of a two-spot model is motivated by our search for stellar differential rotation. Nevertheless, given the nearly sinusoidal shape of the rotational modulation of our light curves with an almost constant phase of their minima and a slow variation of their amplitudes (cf. Fig.~\ref{Fig1}), it is useful to consider an alternative model with a single  spot whose area is evolving in time according to a quadratic law, i.e., $A_{1} (t) = A_{10} + A_{11}(t-t_{0}) + A_{12} (t-t_{0})^{2}$, where $t$ is the time, $t_{0}$ the initial time of each of the individual intervals to be fitted, $A_{10}$ the area at time $t_{0}$, $A_{11}$ and $A_{12}$ coefficients found by fitting the light modulation. We add a time variable longitude-independent flux term, i.e., $F_{0}(t) = F_{00} + F_{01} (t-t_{0}) + F_{02} (t-t_{0})^{2}$, with a similar meaning of the symbols, to take into account a simultaneous variation of the uniformly distributed background of small spots. This model is compared with the two-spot model introduced above to assess the actual improvement obtained by using two spots instead of a single evolving spot when fitting our light curve segments. Both models have ten free parameters, thus allowing us a simpler statistical comparison of their relative performance.

\section{Stellar parameters}
\label{parameters}
The limb-darkening coefficients adopted for our stars are indicated in Table~\ref{table2}, together with the  effective temperature $T_{\rm eff}$ and the gravity $\log g$ assumed to compute them after \citet{ClaretBloemen11}.
{The effective temperature $T_{\rm eff}$ comes  from the spectral type (cf. Table~\ref{table1}) and the \citet{KenyonHartmann95} (hereafter KH95) temperature scale. To evaluate the surface gravity $g$ we need the mass and the radius of the star. The mass is computed from $T_{\rm eff}$, the bolometric luminosity $L_{\rm bol}$, and \citet{Siessetal00} pre-main-sequence evolutionary models. The radius is estimated from $T_{\rm eff}$ and $L_{\rm bol}$. The latter is  estimated from the $I_{\rm c}$ magnitude, the bolometric correction on $I_{\rm c}$, $BC_{I}$, the absorption on $I_{\rm c}$, $A_{I}$,  and the adopted distance to NGC~2264 \citep[760~pc,][]{Sungetal97}. The bolometric correction $BC_{I}$ is estimated from the spectral type and the KH95 tabulation; $A_{I}$ comes from the intrinsic $R-I$, a function of the spectral type (again from KH95), the observed $R-I$, and the \citet{Mathis90} extinction law. 
}
 The metallicity of NGC~2264 is assumed to be $ \rm [Fe/H] = -0.15$ \citep{Kingetal00,Heiteretal14}. The spot contrast is fixed at  the solar value ($c_{\rm s} = 0.677$) because we have no information on spot temperature. 

To apply our MCMC Bayesian approach to the light curve segments, we  assume uniform a priori distributions of the model parameters within the ranges specified in Table~\ref{apriori_params}. The initial and final times $t_{1}$ and $t_{2}$ of each modelled light curve segment are measured from the beginning of its light curve, respectively. The area of the spots $A_{k}$ is expressed as a fraction of the stellar disc, i.e., $a_{k} \equiv A_{k}/\pi R_{*}^{2}$, $R_{*}$ being the radius of the star, while the ranges of the spot longitudes are given as the maximum deviations from their initial values $\Delta \lambda_{k}$, with $k=1,2$. The  parameter values of the starting point of the MCMC are those corresponding to the minimum of the chi square for each light curve segment { (see Appendix~\ref{app1})}. 
\begin{table}
\caption{Stellar parameters and quadratic limb-darkening parameters. }
\begin{tabular}{cccccc}
\hline
 & & & & & \\
CoRoT ID & $T_{\rm eff}$ & $\log g$ & $a_{\rm p}$ & $b_{\rm p}$ & $c_{\rm p}$ \\ 
 & (K) & (cm~s$^{-2}$)  \\
 & &  \\
 \hline
 & &  & & & \\
 223959652 & 6030 & 4.17 &  0.370 &  0.919 &   -0.289 \\
 616849446 & 4470 & 3.84  & 0.258 &    0.812 &   -0.070 \\
 223982407 & 4350 & 4.06 &  0.245  & 0.784 &   -0.029 \\
 223988965 & 4205 & 3.72 & 0.238   &  0.774 &   -0.012 \\ 
  616919771 &  4350 & 3.87 &  0.240  &  0.766 &  -0.006 \\
 & & & & & \\
 \hline
\end{tabular}
\label{table2}
\end{table}

\begin{table*}
\caption{Initial and final times of the segments considered for the MCMC analysis, together with the uniform prior intervals of the two-spot model parameters for each star.  }
\begin{center}
\begin{tabular}{lcccccccc}
\hline
\hline 
Star CoRoT ID & $t_{1}$ & $t_{2}$ & $i$ & $F_{0}$ & $a_{1,2}$ & $\theta_{1,2}$ & $\Delta \lambda_{1,2}$ & $P_{1,2}$ \\
  & (d) & (d) & (deg) & & & (deg) & (deg) & (d) \\
  \hline
  & & & & & & & \\
 223959652 & 32.21202 & 38.50473 &  30, 90 & -0.02, 0.01 & 0.001, 0.3 & 0, 180 & $\pm 15$ & 2.0, 5.0 \\
 616849446 & 23.19344 & 30.77758 & 30, 90 & -0.02, 0.01 & 0.001, 0.268 & 0, 180 & $\pm 15$ & 2.5, 6.0 \\
 223982407 &  30.92017 & 38.36193 & fixed & fixed & 0.001, 0.4 & 0, 180 & $\pm 15$ & 2.0, 3.0 \\
 223988965 & 0.0 & 6.30319 & fixed & fixed & 0.001, 0.3 & 0, 180 & $\pm 15$ & 2.0, 6.0 \\
 616919771 & 0.0 & 12.75062 & 5, 90 & -0.02, 0.01 & 0.001, 0.268 & 0, 180 & $\pm 15$ & 5.0, 15.0 \\
 \hline
\end{tabular}
\label{apriori_params}
\end{center}
\end{table*}

\section{Results}
\label{results}
The autocorrelation functions of our five photometric time series are plotted in Fig.~\ref{autocorrelation}. The ratio of  the second maximum to the first maximum is greater than 0.6 in all the cases, indicating that the light curves of these WTTs are stable enough to allow us a search for differential rotation. The dotted lines indicate the $\pm \sigma$ range, where $\sigma$  is the standard deviation of the autocorrelation function in the case of a  random variable with some degree of autocorrelation according to the so-called large-lag approximation \citep[see Sect.~3.1 in][for detail]{Lanzaetal14}. 

By applying our approach, we selected segments with the best evidence of DR for all our targets. The minimum value of $\Delta {\rm BIC}$ for them was 2.2, providing preliminary evidence for DR in all our cases. { These segments were fitted also with the evolving single spot plus  background model and the ratio $\chi^{2}_{\rm 1s}/\chi^{2}_{2s}$ of the $\chi^{2}$ values obtained with that model and the two-spot model is reported in Table~\ref{table_2s-1s_ratio} together with the relative $\Delta$BIC and its significance  as obtained by the F-statistics, given that both models have the same number of free parameters.  The significance is the probability of obtaining a $\chi^{2}_{\rm 1s}/\chi^{2}_{2s}$  equal to or smaller than the listed value in the case of a random variable. Therefore, values close to the unity are an indication that the two-spot model provides a statistically better fit than the one-evolving spot model. This is always the case, except for 223988865 for which we have comparable $\chi^{2}$ values with both models. Nevertheless,  we shall consider the two-spot model  also for that star. }
\begin{table}
\caption{Statistical comparison between the best fits obtained with the two-spot and the evolving single-spot models with the Levenberg-Marquardt and Nelder-Mead optimizations for the light curve segments selected to search for differential rotation.  }
\begin{center}
\begin{tabular}{lccc}
\hline
\hline 
CoRoT ID & $\chi^{2}_{\rm 1s}/\chi^{2}_{\rm 2s}$ & $\Delta$BIC & Significance \\
\hline
 & & & \\
223959652 &        2.048 &        2.096 &         0.980 \\
616849446 &        3.494 &        4.989 &         1.000 \\
223982407 &        4.073 &        6.147 &         1.000 \\
223988965 &        0.974 &       -0.053 &         0.470 \\
616919771 &        1.377 &        0.755 &         0.917 \\
\hline
\end{tabular}
\label{table_2s-1s_ratio}
\end{center}
\end{table}

The segments  selected with the above procedure were successively fitted with the MCMC algorithm to derive the a posteriori distributions of the spot model parameters and confirm the preliminary results based on the best fits computed with the Levenberg-Marquardt { and Nelder-Mead algorithms (see Appendix~\ref{app1} for a comparison of the respective results). For CoRoT~223982407 and~223988965, it was not possible to achieve convergence of the MCMC for the two-spot model even after refining the starting point of the chain with the Metropolis-Hasting algorithm and running chains of several hundred million points. Only after fixing the inclination of the stellar spin axis $i$ and the uniform flux term $F_{0}$, we were able to obtain convergence as measured by  an $\hat{R}$ statistics smaller than 1.1 (cf. Table~\ref{r_table}) along chains of 150 and 80 million points, respectively. 

The best fits of the light curves corresponding to the minimum chi squares as obtained with the Metropolis-Hasting algorithm are plotted in Figs.~\ref{Fig_best_fits} and \ref{Fig_best_fits_1} for the two-spot model and the evolving single spot model plus background, respectively. The best fits obtained with the two-spot model are statistically better than those obtained with the evolving single spot model plus background as indicated by the significantly lower values of the $\chi^{2}$ in Table~\ref{chi_square_values} (we list in Appendix~\ref{app2} all the best fit parameters with their uncertainties). The difference is clearly visible in the case of  CoRoT  223982407 and, to a lesser extent, also for the other stars. The single spot model can reasonably fit the minima of the light curves, but the shape of the overall light variation is better reproduced with two spots whose longitudes can change from one rotation to the other. This confirms the results obtained with the Levenberg-Marquardt and Nelder-Mead optimizations. }The values of $\chi^{2}_{\rm red}$ are significantly larger than the unity indicating that the variability of the light curves on timescales shorter than the mean rotation period is significantly larger than the standard errors of the individual data points as reported in Table~\ref{table1}. This can be ascribed to several small spots that are not included in our model or to flaring activity of the targets, e.g., in the case of 223988965 (cf. Fig.~\ref{Fig_best_fits}). A similar situation has been discussed by \citet{Lanzaetal14} in  the case of Kepler-30. { To quantify the short-term variability, we smoothed the observed light curves with a sliding box-car filter and computed the standard deviation of the residuals. For timescales from $0.1$ to $0.15$ of the mean rotation periods, that is much shorter than the phase range that can be accurately fitted with a two-spot model,  we find a standard deviation in relative flux units of $\sim 10^{-3}$ for CoRoT~223959652, 223988965, and 616919771, of $ 2.6 \times 10^{-3}$ for CoRoT~616849446, and of   $4.3 \times 10^{-3}$ for CoRoT~223982407. Given that the reduced $\chi^{2}$ listed in  Table~\ref{chi_square_values} have been calculated considering the standard errors in Table~\ref{table1}, their values would become closer to the unity, if we consider the standard deviations on the above time intervals.  
\begin{table}
\caption{Reduced $\chi^{2}$ values obtained with the two-spot and the evolving single-spot  models for our stars, respectively.   The minima of the $\chi^{2}$ are computed with the Metropolis-Hasting algorithm (cf. Appendix~\ref{app1} for details). }
\begin{center}
\begin{tabular}{lrr}
\hline
\hline 
CoRoT ID & $\chi^{2}_{\rm red}$ (2 spots) & $\chi^{2}_{\rm red}$ (1 evol. spot)  \\
\hline
& & \\
223959652 &        28.32 &        54.56 \\
616849446 &        43.34 &     203.45 \\
223982407 &        13.54 &        55.99 \\
223988965 &          8.68  &         8.82 \\
616919771 &        11.98  &        26.46 \\
\hline
\end{tabular}
\label{chi_square_values}
\end{center}
\end{table}

The mean values and the standard deviations of the spot rotation periods as derived with our MCMC approach for our five targets are listed in Table~\ref{diff_rot_table} together with the mean stellar inclination and relative differential rotation $\Delta P/P$, where $\Delta P = | P_{1} - P_{2}|$ and $P = (P_{1} + P_{2})/2$, respectively. The relative differential rotation for CoRoT~223988965 and~616919771 is remarkably larger than for the other three targets, while their inclination  is so low than the reliability of their spot rotation periods is questionable because the spot longitudes are ill-constrained in this case (see Sect.~\ref{discussion}). Furthermore, in the case of CoRoT~223988965, the short-term variability of the light curve makes the fit not as good as in the other cases, while for CoRoT~616919771 the fitted segment covers  one rotation period only. This reduces  the relative longitudinal shift of the spots upon which our method is based and, when combined with a low inclination,  makes our DR measurement uncertain. For these reasons, we do not consider the DR values obtained for those two stars significant. For CoRoT~616919771, we considered also a longer segment of 19.2525 days, but the best fit is not sufficiently good. Its minimum chi square is larger than that of the previously considered shorter segment by a factor of $\sim 4.5$, likely as a consequence of the intrinsic starspot evolution over a longer time interval. The decay of the autocorrelation function is faster for this star than for the other four targets, thus supporting the conclusion that an intrinsic spot evolution together with a low inclination mask any DR signal in its light curve (cf. Fig.~\ref{autocorrelation}, bottom panel). 

In conclusion, we focus on the first three targets whose a posteriori rotation period distributions are plotted in Figs.~\ref{prot_223959652}, \ref{prot_616849446}, and~\ref{prot_223982407}. They have been obtained from MCMCs consisting of 60, 50, and 150 million points, and with an average acceptance fraction of 24.7, 26.7, and 23.7 percent, respectively. The distributions of $P_{1}$ and $P_{2}$ do not overlap, implying that the difference between the two rotation periods is statistically meaningful for all the three stars. {The correlation of $P_{1}$ and $P_{2}$ with the other model parameters is plotted in Fig.~\ref{coupled_distributions} for CoRoT~616849446, considering $15\,000$ representative points extracted from the mid of the MCMC;} similar plots of the joint parameter distributions are obtained also for the other two stars. The rotation periods $P_{1}$ and $P_{2}$ appear to be strongly correlated with the initial spot longitudes $\lambda_{1}$ and $\lambda_{2}$, respectively, but the amplitude of the associated variation is not large enough to hamper the significance of the difference between the two periods. Other remarkable correlations are those between the inclination $i$ and the colatitude of the spots $\theta_{1,2}$ and between the colatitudes and their areas $A_{1,2}$, respectively; they have been discussed in detail by \citet{Lanzaetal14}. 

Finally, we calculated best fits for the entire time series with the Levenberg-Marquardt algorithm by fixing the inclination $i$ and the rotation periods  $P_{1}$ and $P_{2}$ to their mean values as obtained from their a posteriori distributions.  In Figs.~\ref{extended_best_fits} { and~\ref{extended_best_fits1}} we plot both best fits computed for the whole time interval (red solid lines) and for individual segments having the same length of the segment adopted for the MCMC analysis (green solid lines), respectively. When we fit the entire time series, the areas of the spots and their coordinates are held fixed for the whole interval which results in a poor fit; see, for example, the top panel of Fig.~\ref{extended_best_fits}, where the two spots are located on opposite hemispheres during the first part of the time series, but fall on the same hemisphere in the second part because they rotate with $P_{1} \not= P_{2}$, so their longitude separation changes vs. the time. The reproduction of the amplitude variations and times of light minima along this light curve is much better if, in addition to the relative shift of the spots produced by $P_{1} \not= P_{2}$, we allow for a variation of the spot areas and coordinates from one segment of the time series to the next (Fig.~\ref{extended_best_fits1}, solid green line). The same conclusion is found for the other two light curves, although the difference between the two best fits is less pronounced. 
Specifically, the variation of the total area of the spots $A_{\rm tot}= A_{1} + A_{2}$ depends on the spot colatitudes and the variation of the amplitude of the light curve from one interval to the next. In particular, when spot colatitudes are fixed, $A_{\rm tot}$ increases with increasing amplitude, e.g., in the case of CoRoT~223959652, otherwise, it stays approximately constant, e.g., in the case of the  other two stars. 

We conclude that some spot evolution is required to fit our time series because models allowing for a change of the spot parameters, in particular of their areas from one segment to the next, provide significantly better fits of the light modulation. 

In the light of the above results, it is interesting to consider the best fits obtained with the Levenberg-Marquardt and the Nelder-Mead algorithms for the entire light curves also with the evolving single spot model plus background. Although we found that the model with two spots was better in the case of the intervals selected to search for differential rotation, in approximately fifty percent of the  cases the evolving single spot model gave a best fit of comparable quality. For each given interval, which model was better depended on the length of the interval and the details of the light modulation of the star. 
In Fig.~\ref{one_spot_1}, we consider as an example that of CoRoT~223959652 whose light curve is fitted in six individual intervals. The varying light amplitude  is well  reproduced by the evolving single spot model along most of  the modulation, except in the interval considered in our search for the differential rotation where the amplitude of the residuals with the two-spot model is smaller.  The characteristic timescale for the evolution of the spot area  $[(1/A_{1}(t))(dA_{1}/dt)]^{-1}$ as derived from the model free parameters is of the order of 20 days, i.e., comparable with that shown by the variation of the light curve amplitude.  However, in the case of other light curves, the timescales can be significantly smaller as the model tries to reproduce the small asymmetries of the light modulations by a faster evolution of the spot and background term.  

\begin{table*}
\caption{Gelman \& Rubin $\hat{R}-1$ statistics measuring the convergence and mixing of the chains of the different parameters of our two-spot models.}
\begin{center}
\begin{tabular}{lcccccccccc}
\hline
\hline
Star CoRoT ID & $i$ & $F_{0}$ & $a_{1}$ & $\theta_{1}$ & $\lambda_{1}$ & $P_{1}$  &  $a_{2}$ & $\theta_{2}$ & $\lambda_{2}$ & $P_{2}$ \\
\hline
 & & & & & & & & & & \\
223959652 & 0.0652 & 0.0260 & 0.0451 & 0.0657 & 0.0873 & 0.0852 & 0.0733 & 0.0782 & 0.0456 & 0.0295 \\
616849446 & 0.0404 & 0.0745 & 0.0369 & 0.0359 & 0.0313 & 0.0264 & 0.0124 & 0.0186 & 0.0346 & 0.0309 \\ 
223983407 & $-$ & $-$ & 0.0104 & 0.0081 & 0.0182 & 0.0152 & 0.0107 & 0.0085 & 0.0199 & 0.0179 \\
223988965 & $-$ & $-$ & 0.0006 & 0.0007 & 0.0040 & 0.0046 & 0.0035 & 0.0026 & 0.0013 & 0.0015 \\
616919771 & 0.0136 & 0.0075 & 0.0439 & 0.0300 & 0.0138 & 0.0490 & 0.0149 & 0.0152 & 0.0165 & 0.0251 \\
 \hline
\end{tabular}
\label{r_table}
\end{center}
\end{table*}
\begin{table*}
\caption{Inclination of the stellar spin axis, spot rotation periods, and their relative difference as derived through  the MCMC analysis of the light curve segments in Table~\ref{apriori_params}. }
\begin{center}
\begin{tabular}{lcccccccc}
\hline
\hline
Star CoRoT ID & $i$ & $\sigma_{i}$ & $P_{1}$ & $\sigma_{P_{1}}$ & $P_{2}$ & $\sigma_{P_{2}}$ & $\Delta P /P$ & $\sigma_{\Delta P}/P$ \\
& (deg) & (deg) & (d) & (d) & (d) & (d) &  & \\
\hline
 & & & & & \\
223959652 & 82.065 & 0.464 & 3.79487 & 0.00341 & 3.68093 & 0.00158 & 0.0305 & $8.45 \times 10^{-4}$ \\ 
616849446 & 68.704 & 0.462 & 4.20364 & 0.00458 & 4.41359 & 0.00314 & 0.0487 & $ 6.27 \times 10^{-4}$ \\
223982407 & 51.702 & ($i$ fixed) & 2.58431 & 0.00198 & 2.63194 & 0.00200 & 0.0183 & $5.57 \times 10^{-4}$ \\
223988965 & 15.0 & ($i$ fixed) & 2.90597 & 0.01927 & 5.04884 & 0.06372 & 0.5388 & 0.01674 \\
616919771 & 13.421 & 0.574 & 6.51568 & 0.0389 & 10.49925 & 0.01269 & 0.4682 & 0.00481 \\
\hline
\end{tabular}
\label{diff_rot_table}
\end{center}
\end{table*}

   \begin{figure}
   \centerline{
   \includegraphics[width=8cm,angle=0]{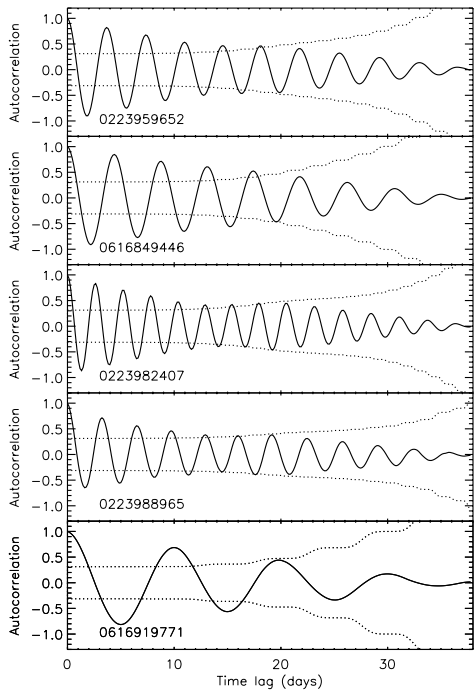}} 
   \caption{Autocorrelation functions of our target light curves as labelled. The dotted lines indicate the $\pm \sigma$ interval (see the text). }
              \label{autocorrelation}%
    \end{figure}
%
 

   \begin{figure}
   \centerline{
   \includegraphics[width=8cm,angle=0]{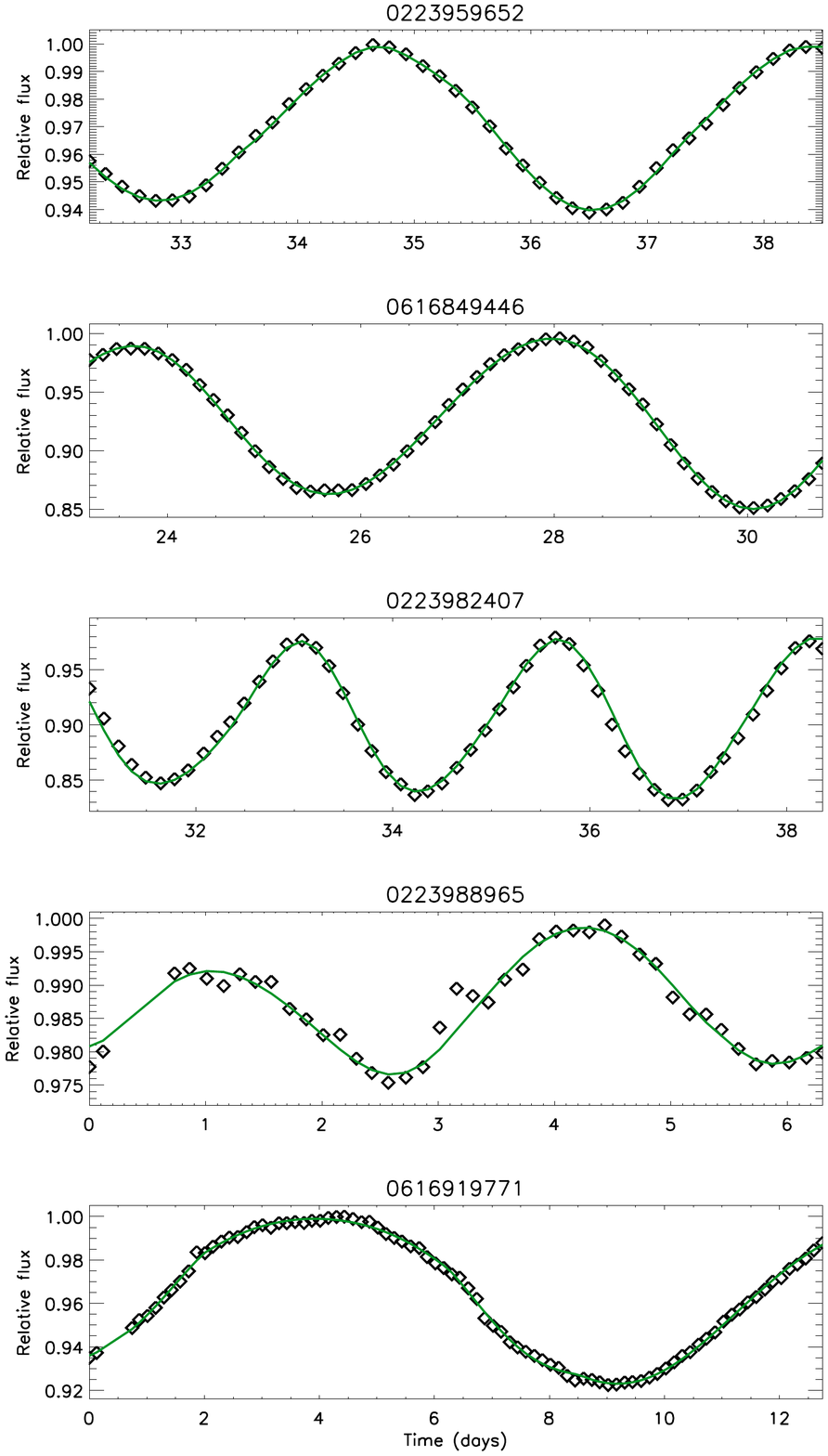}} 
   \caption{Best fits of  the considered segments of our light curves (open diamonds) obtained with  the two-spot model (solid green line) and the Metropolis-Hasting algorithm as labelled. 
   The increase in flux around 3.2 days in the light curve of CoRoT~223988965 is associated with an X-ray flare. }
              \label{Fig_best_fits}%
    \end{figure}
%
 

   \begin{figure}
   \centerline{
   \includegraphics[width=8cm,angle=0]{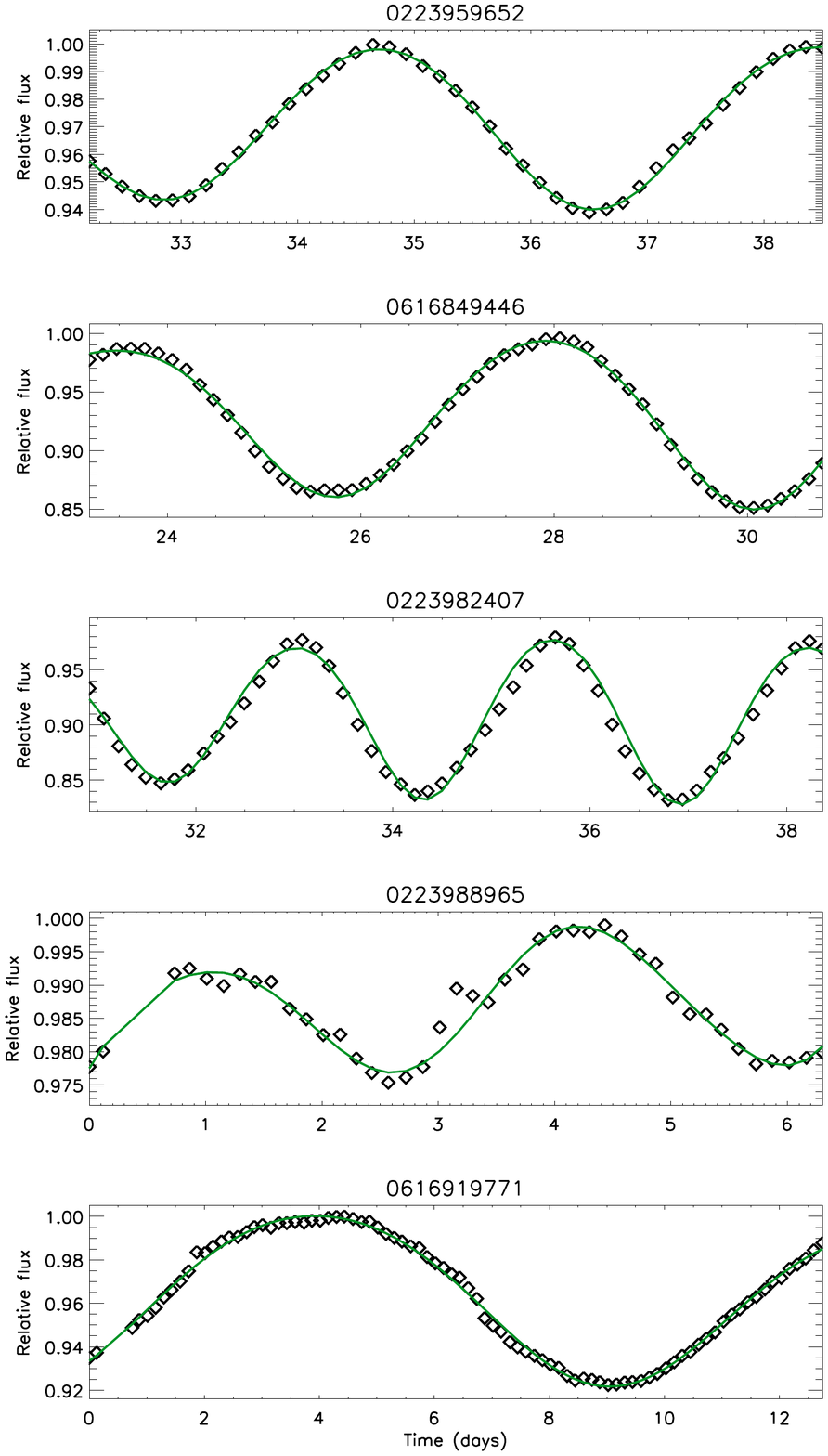}} 
   \caption{Best fits of  the considered segments of our light curves (open diamonds) obtained with  the evolving single-spot model  (solid green line) and the Metropolis-Hasting algorithm as labelled. 
   The increase in flux around 3.2 days in the light curve of CoRoT~223988965 is associated with an X-ray flare. }
              \label{Fig_best_fits_1}%
    \end{figure}
%
 

   \begin{figure}
   \centerline{
   \includegraphics[width=8cm,angle=0]{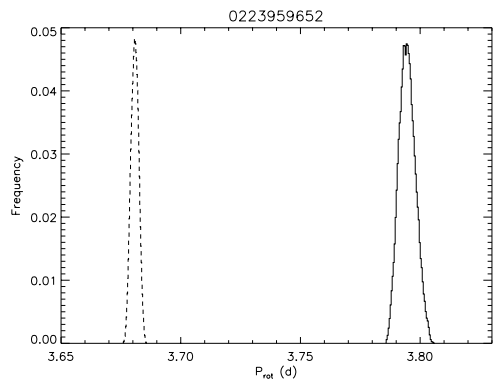}} 
   \caption{A posteriori marginal distributions of the rotation periods of the two spots as derived from the MCMC analysis for CoRoT 223959652. The solid line refers to the distribution of the rotation period of the first spot, the dashed line to that of the second. }
              \label{prot_223959652}%
    \end{figure}
%
 

   \begin{figure}
   \centerline{
   \includegraphics[width=8cm,angle=0]{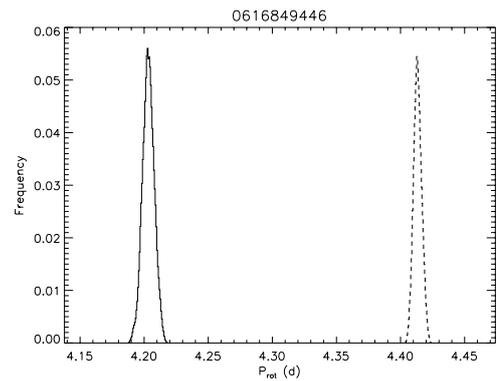}} 
   \caption{Same as Fig.~\ref{prot_223959652} for CoRoT 616849446. }
              \label{prot_616849446}%
    \end{figure}
%
 

   \begin{figure}
   \centerline{
   \includegraphics[width=8cm,angle=0]{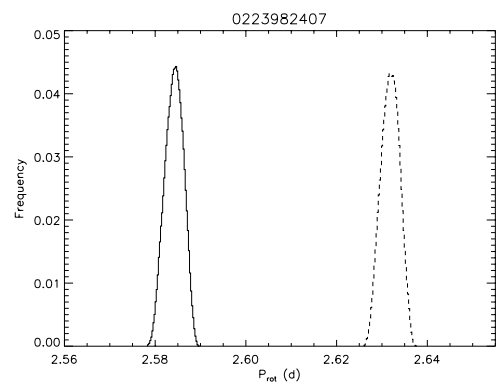}} 
   \caption{Same as Fig.~\ref{prot_223959652} for CoRoT 223982407.  }
              \label{prot_223982407}%
    \end{figure}
%
 
 
   \begin{figure}
   \centerline{
   \includegraphics[width=10cm,height=9cm, angle=90]{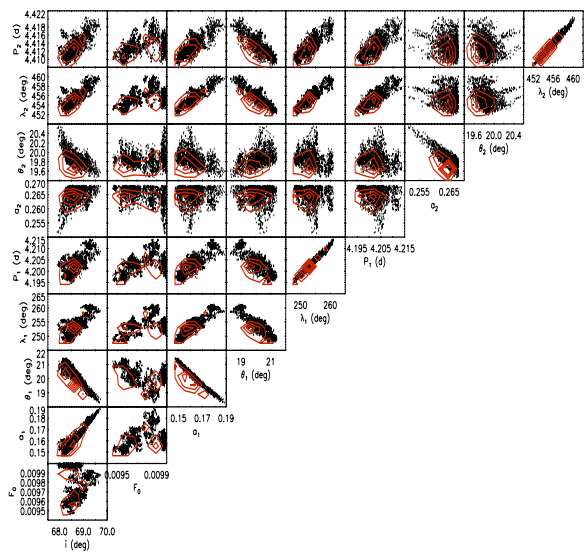}} 
   \caption{A posteriori joint distributions  of couples of model parameters in the two-spot model of CoroT 616849446 as derived from our MCMC analysis. Density isocountours as obtained with the {\tt CONTOUR} procedure of IDL are superposed (red solid lines). }
              \label{coupled_distributions}%
    \end{figure}
%
 
 
   \begin{figure}
   \centerline{
   \includegraphics[width=9cm,height=10cm,angle=0]{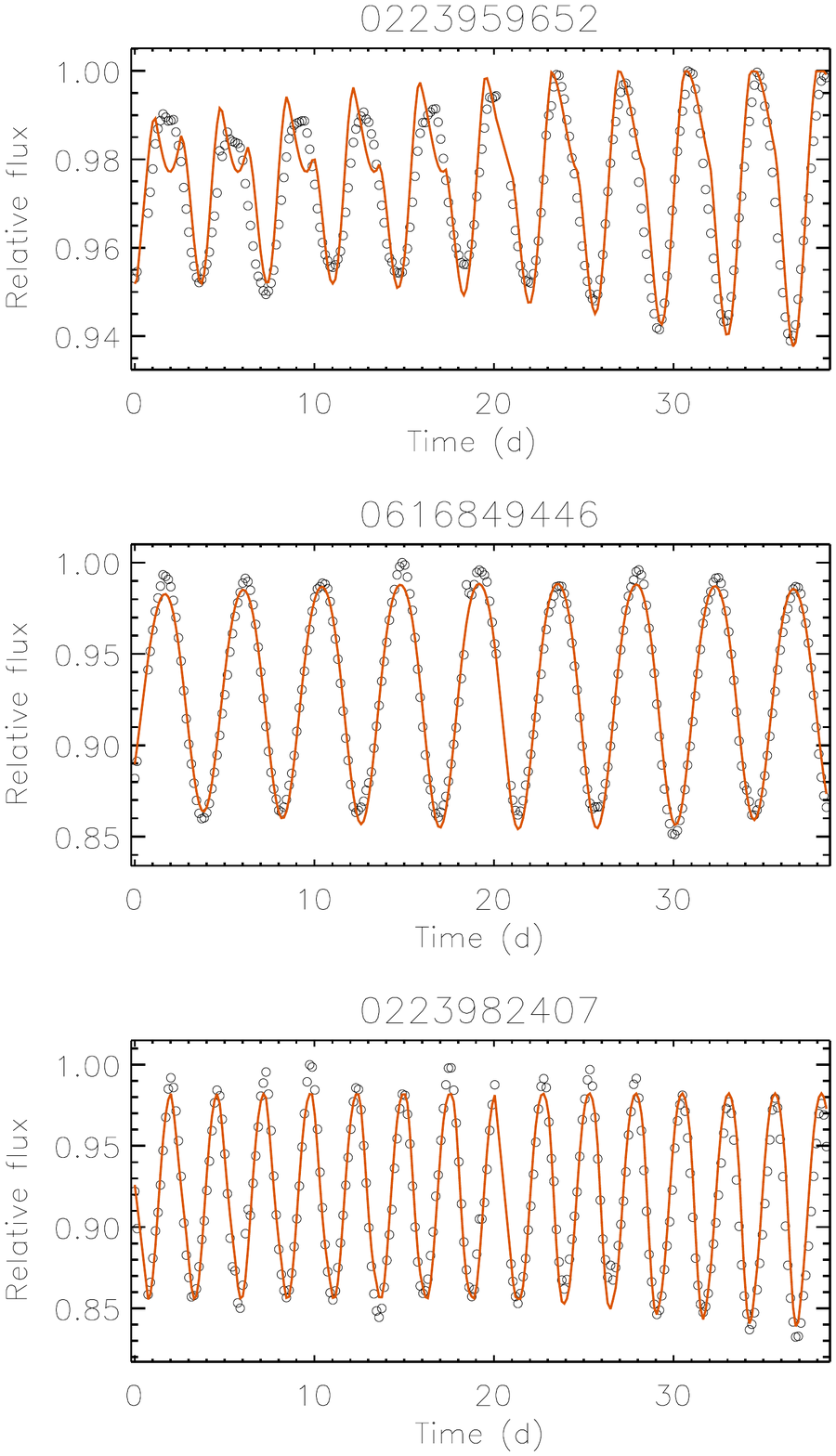}} 
   \caption{Entire light curve best fits with the inclination and the rotation periods of the two spots held fixed at their mean values as derived from the a posteriori distributions computed with the MCMC.   The black open circles indicate the observations, the  orange solid lines  the best fits to the entire light curves (see text).} 
                \label{extended_best_fits}%
    \end{figure}
%
 

   \begin{figure}
   \centerline{
   \includegraphics[width=9cm,height=10cm,angle=0]{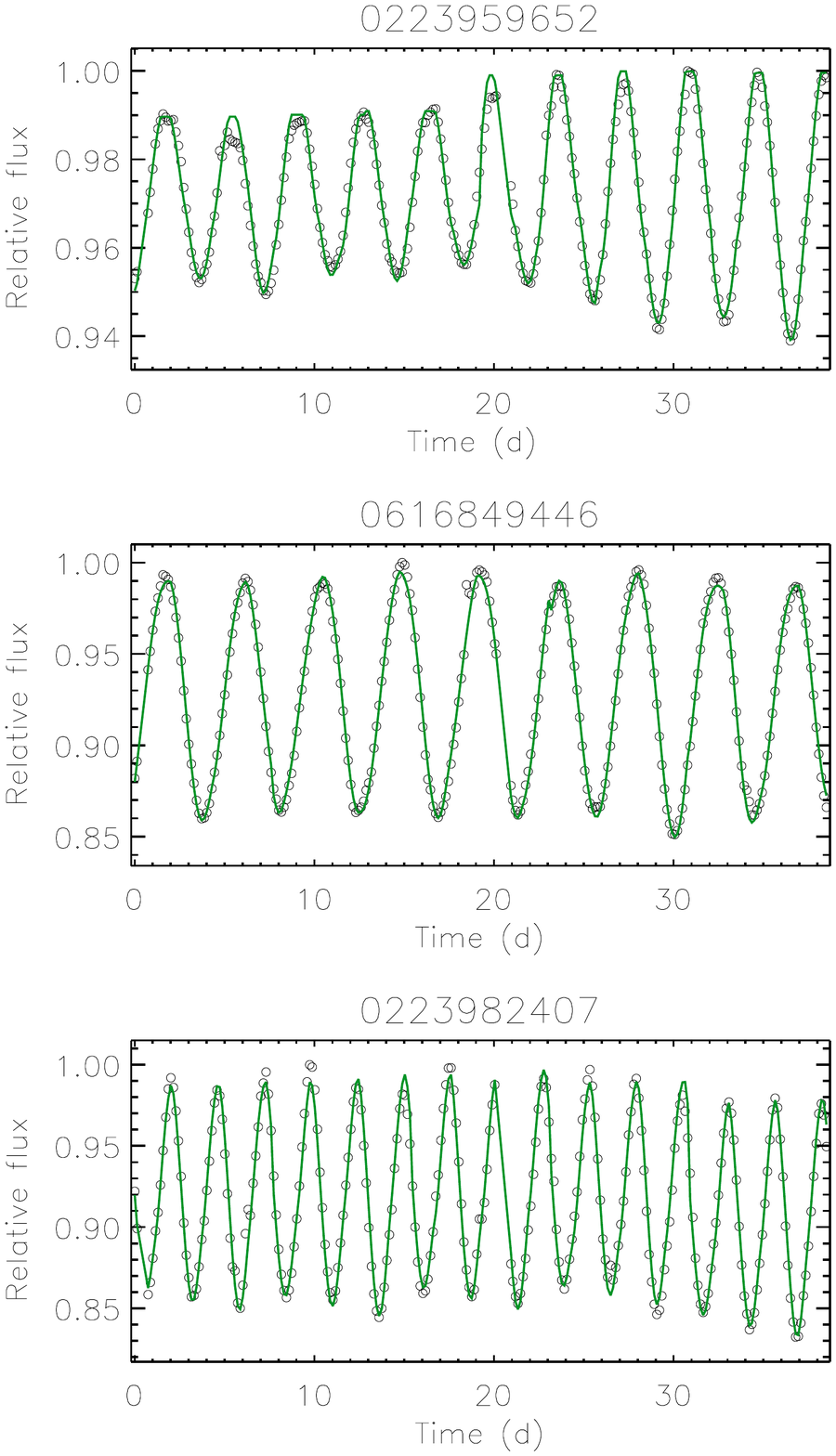}} 
   \caption{
   As Fig.~\ref{extended_best_fits}, but plotting the best fits for the individual light curve segments (green lines) instead of the entire light curves (see text).}
              \label{extended_best_fits1}%
    \end{figure}
%
 

   \begin{figure}
   \centerline{
   \includegraphics[width=9cm,angle=90]{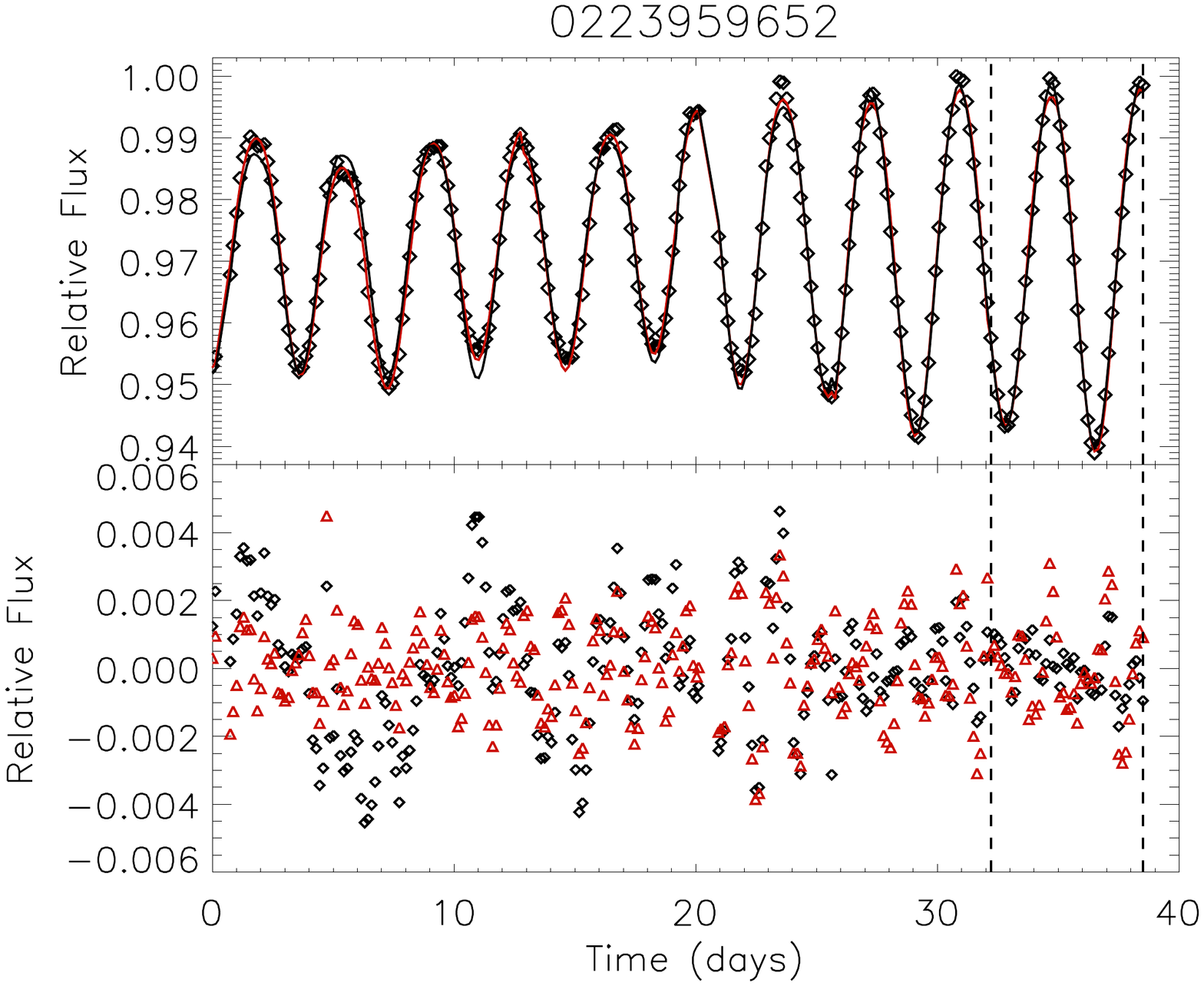}} 
   \caption{Upper panel: The entire light curve of CoRoT~223959652  (open diamonds) together with the best fits obtained by fitting individual intervals of the same length of that used to search for differential rotation (indicated by the dashed vertical lines) with the two-spot model (black solid line) and the evolving single spot model  (red solid line). Lower panel: the residuals of the two best fit models;  those of the two spot model are plotted as black diamonds, while those of the evolving single spot model as red diamonds.    }
              \label{one_spot_1}%
    \end{figure}

\section{Discussion and conclusions}
\label{discussion}

{ We have compared best fits obtained with a two-spot model and an evolving single spot plus background in the case of TTSs whose light curves appear to be dominated by photospheric brightness inhomogeneities that vary slowly from one rotation to the next. We found time intervals during which the two-spot model with $P_{1} \not= P_{2}$ is statistically better than two rigidly rotating spots or the evolving single spot model and focussed on those intervals to estimate the amplitude of stellar differential rotation.  The Nelder-Mead algorithm improved the best fits obtained with the Levenberg-Marquardt method  in a fraction of cases ranging from 3 to 10 percent, depending on the specific light curve,  when we adopted a significance threshold of 95 percent,  but never changed the value of the $\Delta$BIC to such an extent that the selection of the intervals with evidence of differential rotation was affected. 

The final estimate of the differential rotation was obtained for those intervals by means of an MCMC approach that included a Metropolis-Hasting optimization of the best fit parameters. Therefore, we were able to further improve our solutions and reduce the $\chi^{2}$ in comparison to those found with the two deterministic methods considered in the previous stage of interval selection (cf. Appendix~\ref{app1} for more details). 

MCMC was performed by considering the rotation periods of the spots as  independent of their latitudes, respectively. As a matter of fact, one could have adopted a parametric law of the form $P(\theta) = P_{\rm eq}/(1-k \cos^{2} \theta)$, where $P(\theta)$ is the rotation period of a spot at colatitude $\theta$ and $P_{\rm eq}$ the rotation period at the equator of the star. We did experiments with this law and found that it makes the convergence of the MCMC chains worse, leading  to non-convergent series after $150\times 10^{6}$ and $300 \times 10^{6}$ steps for CoRoT~223959652 and 616849446, respectively. For CoRoT~223982407 we got a worse convergence after $150 \times 10^{6}$ steps with a maximum value of the Gelman-Rubin parameter $\hat{R} = 1.189$. The equatorial period $P_{\rm eq}$ was found to be $2.731 \pm 0.012$~days, while  the parameter $k = -0.0950 \pm 0.0095$ indicated an amplitude of the differential rotation $\sim 4.5$ times larger than that estimated from the relative difference of the two periods when they were not tied to the spot colatitudes. This result is in line with that of \citet{Croll06} who adopted a similar model with the above parameterized differential rotation law in the case of $\epsilon$~Eri. The 68 percent mean likelihood credible interval in that case was $0.02 \la k \la 0.2$ (cf. his Fig.~2, lower right panel), while the single spot rotation periods were much better defined (cf. his Fig.~3). We conclude that the use of a parametric differential rotation law, although physically motivated, amplifies the dependences among model parameters, in particular between spot periods and colatitudes, thus making the convergence generally worse. Therefore, adopting spot periods as independent parameters is better also in view of the very limited information on spot latitude that can be extracted from wide-band light curves. 
}

{ To further understand  the results obtained with the MCMC method in the case of CoRoT~223988965 and~616919771, we consider the critical role played by the inclination of the stellar spin axis~$i$ in determining the accuracy of the spot longitudes as illustrated by Fig.~\ref{lc_minimum}.} When $i$ is low, a high-latitude spot is always in view and its light modulation has a shallow minimum that makes  its longitude uncertain  (top panel). On the other hand, when $i$ is high, the spot is visible only for part of a rotation and the minimum of its light modulation is narrower and deeper making the determination of its longitude more accurate because the time of its transit across the central meridian of the stellar disc is better defined (lower panel). Since an accurate determination of the spot longitudes is at the base of the measurement of their rotation periods, the values of $P_{1}$ and $P_{2}$ are significantly more accurate in the case of the first three targets for which $i \geq 50^{\circ}$, while they are ill defined in the case of the other two targets for which $i \leq 15^{\circ}$. Therefore, we do not trust their spot rotation periods. 

Differential rotation was measured through Doppler Imaging techniques in four WTTs, i.e., V410~Tau \citep{Skellyetal10}, LkCa~4 \citep{Donatietal14}, V819~Tau, and V830~Tau \citep{Donatietal15} finding values in agreement with those derived from the long-term variations of their photometric periods \citep[e.g.,][]{Grankinetal08} and ranging from nearly solid-body rotation up to $\Delta \Omega / \Omega = 0.0055 \pm 0.002$ that is a factor of $3-10$ times smaller than the present estimates for our three targets. 

To explain this discrepancy, we suggest that an intrinsic starspot evolution is affecting our measurements. An illustrative case is displayed in Fig.~\ref{two_spots_sketch}, where we show how our simple two-spot model ($M_{1}$ and $M_{2}$) fits a more complex real spot configuration consisting of three spots ($S_{j}$, with $j=1,2,3$). Consider the case when spot $S_{1}$ stays fixed during the fitted time span, while the two close spots $S_{2}$ \&  $S_{3}$ evolve is such a way that the area of  spot $S_{2}$ decreases while that of spot $S_{3}$ increases with their total area $A_{S_{2}} + A_{S_{3}} $ remaining approximately constant. Our two-spot model can provide a very good fit of the light curve corresponding to this case, provided that the longitude of the model spot $M_{2}$ is steadily increased during the time span of the observations to reproduce the change in the active region $S_{2}$ \& $S_{3}$ whose photocentre is progressively shifted towards $S_{3}$ by the intrinsic spot evolution. In other words, an intrinsic spot evolution can be fitted by a relative drift of the two model spots $M_{1}$ and $M_{2}$ yielding a difference in their rotation periods. 
Indeed, we have found  evidence for intrinsic spot evolution, e.g., by comparing the different models  in Figs.~\ref{extended_best_fits} and~\ref{extended_best_fits1}. 

From a general point of view, our  spot rotation periods can be combined to yield a spot evolutionary time $t_{\rm spot}$  defined as
\begin{equation}
\frac{1}{t_{\rm spot}} = \left| \frac{1}{P_{1}} - \frac{1}{P_{2}} \right|;  
\end{equation}
$t_{\rm spot}$ can be regarded as a combination of the timescales of intrinsic spot evolution and of longitudinal shear associated with differential rotation. {If there were no spot evolution, $t_{\rm spot}$ would provide a measurement of the characteristic shear timescale associated with differential rotation, more precisely $t_{\rm spot} = 1/\Delta \Omega$, where $\Delta \Omega$ is the difference in the angular velocity of the two starspots.} For our three targets, $t_{\rm spot}$ ranges between $\sim 20$ and $\sim 50$ mean rotation periods. In the case of WTTs, the very small amplitude of the differential rotation makes the contribution of the intrinsic spot evolution to $t_{\rm spot}$ significant, in spite of their remarkably stable light curves as indicated by the slow decay of their autocorrelation functions (cf. Fig.~\ref{autocorrelation}). { Note that our evolving single spot model gives a timescale of spot evolution between 1.8 and 6.6 rotation periods for the three stars with a meaningful period difference. Such  short timescales are a consequence of using only one spot to fit light modulations that are not perfectly symmetric about their minima. In other words, to improve the best fit, the model adopts a fast evolution of the single spot to reproduce non-perfectly symmetric light minima that would be better fitted with two (or more) spots. On the other hand, introducing spot evolution directly into our two-spot model would imply an increase in the number of free parameters and in their correlations putting at risk the MCMC convergence and the possibility of finding a meaningful Bayesian model fitting our light curves.} 

{ We conclude that}, our estimate of $t_{\rm spot}$ primarily provides a measurement of the intrinsic evolutionary timescale of the spot pattern with the contribution of the differential rotation being  secondary, given that the associated {shear} timescale $1/\Delta \Omega$ is about $3-10$ times longer {according to the Doppler Imaging measurements on similar stars. The evolution timescales of sunspots and starspots in different active stars have recently been discussed by \citet{BradshowHartigan14} finding that the mean spot lifetime is remarkably shorter in main-sequence stars than in the three TTVs analysed in this work \citep[see also][]{Hussain02}.} 

   \begin{figure}
   \centerline{
   \includegraphics[height=7cm,angle=0]{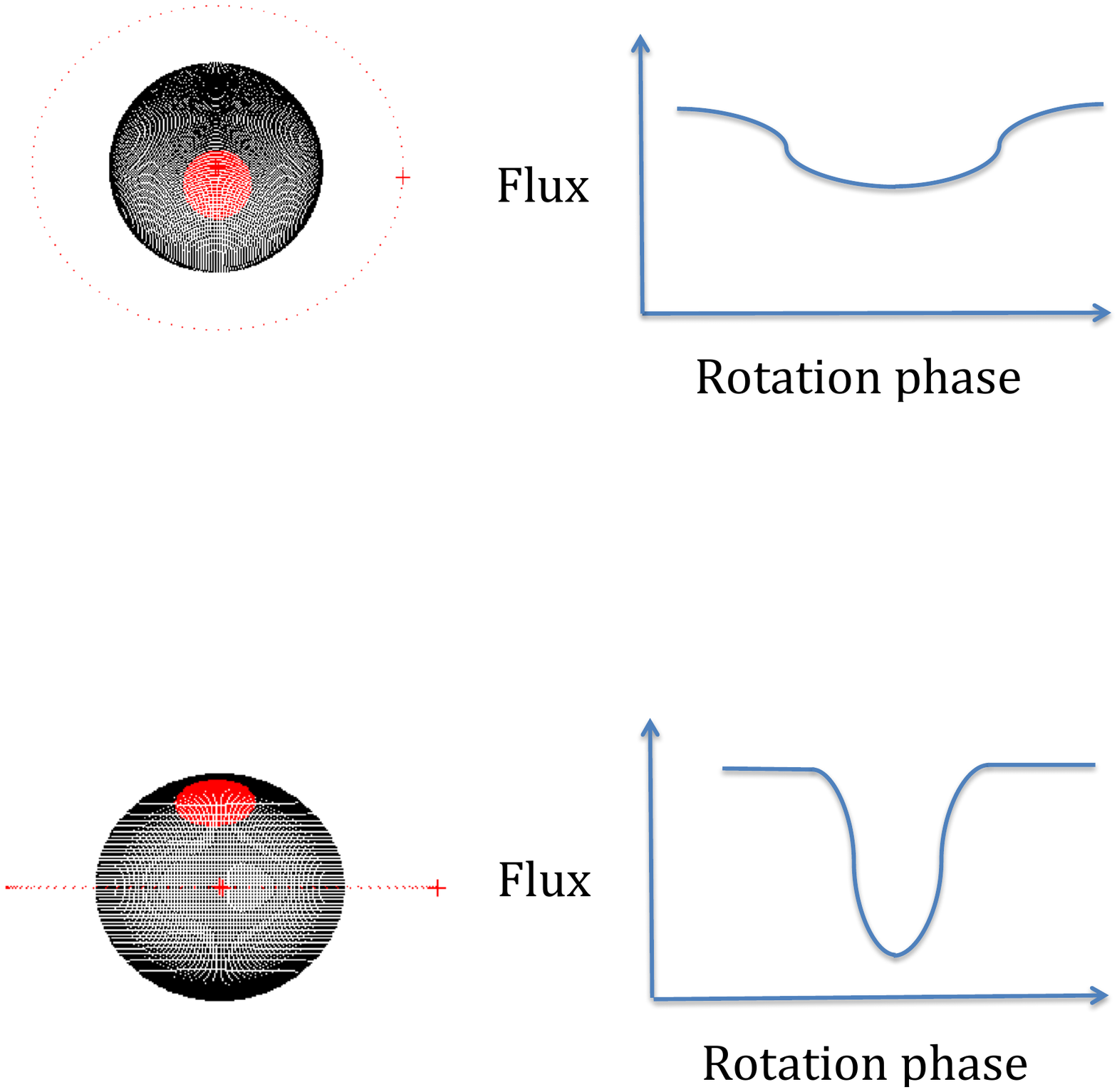}} 
   \caption{Illustration of the effect of the stellar spin axis inclination on the shape of the light modulation produced by a single spot. The case of a high-latitude spot in a low-inclination star is shown in the top panel, while that of a star with a high inclination in the bottom panel. The spot is plotted in red over the disc of the star, while the dotted circle indicates the equatorial plane of the star.}
              \label{lc_minimum}%
    \end{figure}
%
 

   \begin{figure}
   \centerline{
   \includegraphics[height=7cm,angle=0]{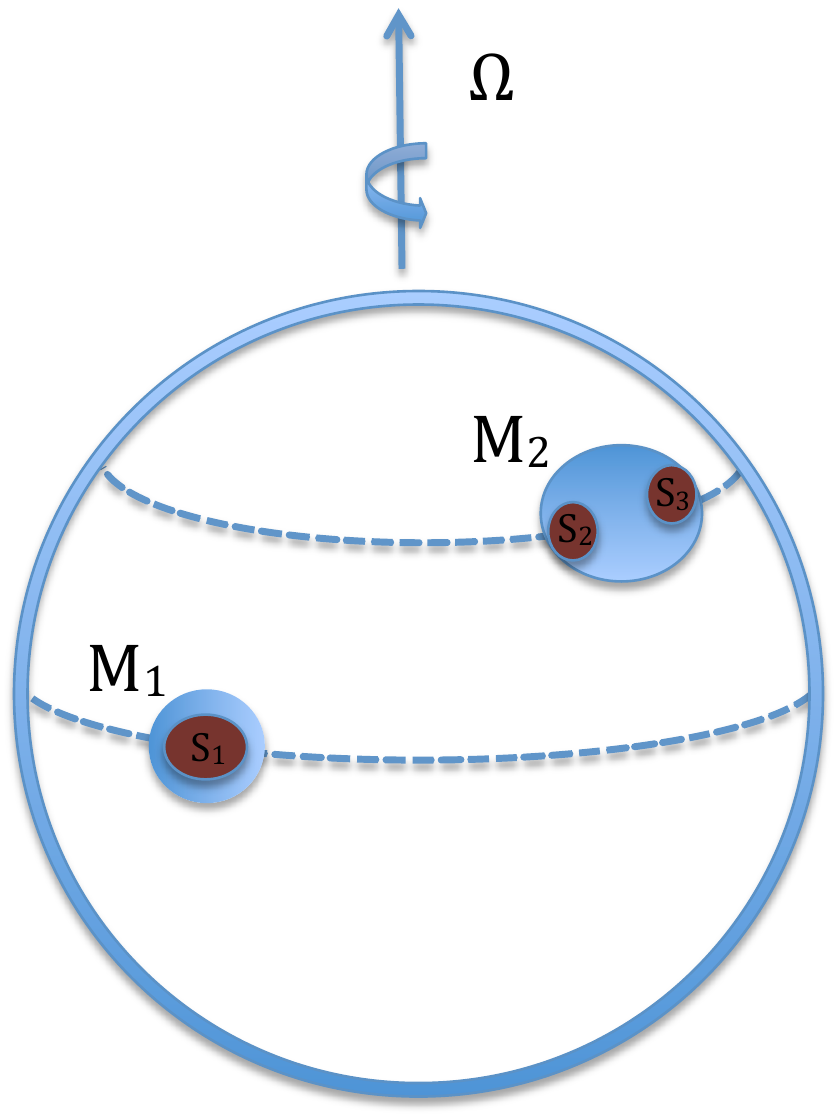}} 
   \caption{Sketch of the representation of a more complex starspot configuration ($S_{1}$, $S_{2}$, and $S_{3}$) by means of only two spots $M_{1}$ and $M_{2}$ (see the text). }
              \label{two_spots_sketch}%
    \end{figure}
%
 

\begin{acknowledgements}
The authors are grateful to an anonymous referee for a careful reading of the manuscript and several stimulating comments that helped to improve this work. 
AFL and SM are grateful to Drs. J.~Bouvier, A.~Frasca, A.~Klutsch, A.~C.~Lanzafame, and E.~Moraux for interesting discussions. 
The authors acknowledge support  through the {\it PRIN-INAF 2012}  funding scheme of the National Institute for Astrophysics (INAF) of the Italian Ministry of Education, University and Research. This work is based on data obtained by the CoRoT Space Experiment and made available through the CoRoT Archive at the IAS Operation Center whose availability is grateful acknowledged. CoRoT is a space mission of the French Space Agency CNES in association with French CNRS Laboratories and partecipation of Austria, Belgium, Brazil, Germany, Spain, and the European Space Agency. 
\end{acknowledgements}

\appendix
\section{Levenberg-Marquardt and Nelder-Mead vs. Metropolis-Hasting optimization}
\label{app1}

{ The $\chi^{2}$ minima of our two-spot models found by means of the Metropolis-Hasting algorithm by running chains starting from the provisional minima found by the Levenberg-Marquardt or Nelder-Mead methods are always better than those provided by those deterministic methods. We run the Metropolis-Hasting optimization several times considering individual chains of $10-15$ milion points that restart from the minimum found by the previous chain until it is not possible to optimize the $\chi^{2}$ further. We set a criterion of acceptance for successive steps  that discards all points where the $\chi^{2}$ increases by more than 1 percent with respect to the assumed minimum in the chain. Intuitively, one may think that this remarkably limits the capability of the algorithm to explore regions of the parameter space far away from its starting point, but we find that this is not the case because the chains move most of the time along the paths determined by the parameter correlations that connect distant points in the parameter space, while the adopted criterion reduces the waste of time in exploring vast domains  where the $\chi^{2}$ is unacceptably large and the fit relative poor \citep[cf.][]{Croll06,Lanzaetal14}. The random  parameter variations from one step to the next allow Metropolis-Hasting to explore extensive domains in the parameter space when tens of millions of steps are considered as in the present case, thus finding the best minimum even if it is very far from the starting point. 

To quantitatively illustrate the improvement provided by the Metropolis-Hasting method over the two other deterministic optimizations, we compare the reduced $\chi^{2}$ values and the parameters at the respective minimum points for our target stars in Table~\ref{table_opt_comp}. The minima found with the deterministic search methods are indicated with LN, while those obtained with the Metropolis-Hasting method are indicated with MH in the second column. The other columns list the reduced $\chi^{2}$ and the other model parameters as in previous similar tables. Note that the inclination was fixed to $60^{0}$ in the case of the LN optimization, while it was varied in the search with MH. The performance of the MH algorithm is always better than those of the two deterministic search methods with a significant variation of the parameters of the minima found with MH with respect to those found with LN. Only the longitudes of the spots and their rotation periods are similar as expected for intervals that were selected to search for differential rotation.  }

\begin{table*}
\caption{Comparison of the $\chi^{2}$ minima obtained with the Levenberg-Marquardt and Nelder-Mead methods with those obtained with the Metropolis-Hasting method.}
\begin{center}
\begin{tabular}{lcccccccccccc}
\hline
\hline
CoRoT ID & Method & $\chi^{2}_{\rm red}$ & $i$ & $F_{0}$ & $a_{1}$ & $\theta_{1}$ & $\lambda_{1}$ & $P_{1}$  &  $a_{2}$ & $\theta_{2}$ & $\lambda_{2}$ & $P_{2}$  \\
& & & (deg) & & & (deg) & (deg) & (day) & & (deg) & (deg) & (day)  \\
\hline
 & & & & & & & & & & \\
223959652 & LN & 28.67 &  60.0 & -0.001394 & 0.0446 & 45.44 & 157.38 & 3.7734 & 0.0536 & 38.31 & 344.59 & 3.6776\\
223959652 & MH & 28.32 & 82.61 & -0.001076 & 0.1431 & 14.41 & 183.64 & 3.7971 & 0.2995 & 9.57 & 355.81 & 3.6811 \\
616849446 & LN &  45.15 & 60.0    & 0.032401 & 0.1811 & 24.62 & 109.55 & 4.4372 & 0.1762 & 16.94 & 261.16 & 4.2051 \\
616849446 & MH & 43.34 & 68.68  & 0.009971  & 0.2680 & 19.68 & 94.19 & 4.4112 & 0.1600 & 20.29 & 251.59 & 4.2005\\
223983407 & LN &  15.21 & 60.0 & -0.020268 & 0.1160 & 38.15 & 181.15 & 2.5732 & 0.1419 & 40.15 & 5.5143 & 2.6185 \\
223983407 & MH & 13.54 & 51.70 & -0.017169 & 0.1172 & 39.36 & 208.02 & 2.5840 & 0.1176 & 50.73 & 34.53 & 2.6319\\
223988965 & LN & 8.99 & 60.0 & 0.006453 & 0.0105 & 64.47 & 233.94 & 5.1133 & 0.0585 & 11.77 & 15.75 & 2.9255 \\
223988965 & MH & 8.68 & 15.0 & 0.008451 & 0.0770 & 87.78 & 231.03 & 5.0562 & 0.0430 & 48.91& 13.13 & 2.9084 \\
616919771 & LN & 19.75 & 60.0 & 0.002054 & 0.0446 & 17.65 & 271.84 & 5.2989 & 0.1391 & 70.66 & 42.75 & 10.3389\\
616919771 & MH & 11.98 & 12.79 & 0.009989 & 0.2617 & 97.08 & 312.53 & 6.4979 & 0.2672 & 69.60 & 45.21 & 10.4917\\ 
 \hline
\end{tabular}
\label{table_opt_comp}
\end{center}
\end{table*}
\section{Best fit parameters and their uncertainty ranges}
\label{app2}

{ We list in Tables~\ref{table_best_fit1}, \ref{table_best_fit2}, and~\ref{table_best_fit3} the parameters corresponding to the two-spot model best fits of the individual light curve segments adopted to search for differential rotation for each of our target, respectively. The uncertainty ranges of the parameters are computed from their marginalized a posteriori distributions as obtained with our MCMC approach and correspond to a probability of 68.2 percent. Note that the uncertainties of the rotation periods in Table~\ref{diff_rot_table} are the standard deviations of the distributions. They are always close to half the ranges listed here because the distribution of $P_{1}$ and $P_{2}$ are close to normal ones. However, this is not the case for all the other distributions, therefore we adopted the above prescription to compute the uncertainty ranges. For each of the model parameter, we add further subscripts $1$ and $2$ to indicated the lower and upper limits of its uncertainty range. For example, $a_{11}$ and $a_{12}$, indicate the uncertainty range of the parameter $a_{1}$ whose best fit value is listed for convenience between the two extrema. } 
\begin{table*}
\caption{Parameters of the best fits of the light curve segments obtained with the Metropolis-Hasting method and their uncertainty ranges (see text). For CoRoT~223982407 and~223988965 the inclination $i$ and the background flux $F_{0}$ were held fixed while running their MCMCs.}
\begin{center}
\begin{tabular}{lcccccccc}
\hline
\hline
CoRoT ID & $\chi^{2}_{\rm red}$ & $i_{1}$ & $i$ & $i_{2}$ & $F_{01}$ & $F_{0}$ & $F_{02}$   \\
\hline
 223959652 &  28.32 &    81.88 &    82.61 &    82.85  & -0.001144 &  -0.001076 &  -0.001014    \\
 616849446 &  43.34 &    68.20 &    68.68 &    69.13  &  0.009810 &   0.009971 &   0.010000    \\
 223982407 &  13.54 &    51.70 &    51.70 &    51.70  & -0.017169 &  -0.017169 &  -0.017169     \\
 223988965 &   8.68 &    15.00 &    15.00 &    15.00  &  0.008451 &   0.008451 &   0.008451   \\
 616919771 &  11.98 &    12.64 &    12.79 &    13.55  &  0.009684 &   0.009989 &   0.010000     \\
  \hline
\end{tabular}
\label{table_best_fit1}
\end{center}
\end{table*}
\begin{table*}
\caption{Parameters of the best fits of the light curve segments obtained with the Metropolis-Hasting method and their uncertainty ranges (see text). Continued from Table~\ref{table_best_fit1}.}
\begin{center}
\begin{tabular}{lcccccccccccc}
\hline
\hline
CoRoT ID & $a_{11}$ &  $a_{1}$ & $a_{12}$ & $\theta_{11}$ & $\theta_{1}$ & $\theta_{12}$ & $\lambda_{11}$ & $\lambda_{1}$ & $\lambda_{12}$ & $P_{11}$ & $P_{1}$ &  $P_{12}$    \\
 & & & & (deg) & (deg) & (deg) & (deg)  & (deg) & (deg) & (day) & (day) & (day) \\
\hline
223959652 &  0.1289 &   0.1431 &   0.1550  & 13.81 &   14.41 &   15.70  &  179.82 &   183.64 &   191.89 &   3.7931 &   3.7971 &   3.8061     \\
616849446 &  0.2633 &   0.2680 &   0.2680 &   19.30 &    19.68 &    19.92  &   88.66 &    94.19 &    96.72 &   4.4036 &   4.4112 &   4.4150 \\
 223982407 &  0.1112 &   0.1172 &   0.1190 &  38.87 &   39.36 &   39.98  &  202.02 &   208.02 &   212.16 &   2.5815 &   2.5840 &   2.5859     \\
 223988965 &  0.0718 &   0.0770 &   0.0820 &  87.15 &    87.78 &    88.28  &  227.84 &   231.03 &   235.71 &   5.0000 &   5.0562 &   5.1428 \\
 616919771 &  0.2051 &   0.2617 &   0.2680 &  96.42 &   97.08 &   98.53  &  310.52 &   312.53 &   313.87 &   6.4329 &   6.4979 &   6.5352   \\ 
   \hline
\end{tabular}
\label{table_best_fit2}
\end{center}
\end{table*}
\begin{table*}
\caption{Parameters of the best fits of the light curve segments obtained with the Metropolis-Hasting method and their uncertainty ranges (see text). Continued from Table~\ref{table_best_fit2}.}
\begin{center}
\begin{tabular}{lcccccccccccc}
\hline
\hline
CoRoT ID & $a_{21}$ & $a_{2}$ & $a_{22}$ & $\theta_{21}$ & $\theta_{2}$ & $\theta_{22}$ & $\lambda_{21}$ & $\lambda_{2}$ & $\lambda_{22}$ & $P_{21}$ & $P_{2}$ & $P_{22}$ \\
& & & & (deg) & (deg) & (deg) & (deg) & (deg) & (deg) & (day) & (day) & (day) \\ 
\hline
223959652 &   0.2642 &   0.2995 &   0.3000 &   9.44 &     9.57 &    10.49  &  354.24 &   355.81 &   358.45 &   3.6795 &   3.6811 &   3.6828 \\
  616849446 & 0.1467 &   0.1600 &   0.1697 &  19.53 &   20.29 &   21.38  &  242.24 &   251.59 &   255.65 &   4.1878 &   4.2005 &   4.2057    \\
 223982407 &  0.1160 &   0.1176 &   0.1207 &   48.72 &    50.73 &    51.44  &   29.80 &    34.53 &    38.49 &   2.6297 &   2.6319 &   2.6339 \\
  223988965 &  0.0424 &   0.0430 &   0.0437 & 48.53 &   48.91 &   49.36  &    9.96 &    13.13 &    17.24 &   2.8903 &   2.9084 &   2.9316     \\
616919771 &     0.2447 &   0.2672 &   0.2680 &  68.36 &    69.60 &    70.19  &   44.80 &    45.21 &    45.48 &  10.4529 &  10.4917 &  10.5059 \\
  \hline
\end{tabular}
\label{table_best_fit3}
\end{center}
\end{table*}

\end{document}